\documentclass[aps,prd,preprint,superscriptaddress]{revtex4-1}

\usepackage{graphicx}
\usepackage{booktabs}
\usepackage{hyperref}
\usepackage{amssymb,amsmath,amsfonts}
\usepackage{color}
\usepackage[utf8]{inputenc}
\usepackage{subfig}

\begin{document}

\title{Conundrum for the free energy of a holonomous gluonic plasma at cubic order}

\author{Christiaan P. Korthals Altes}
\email{chrisaltes@gmail.com}
\affiliation{Centre Physique Theorique, Case 907, Campus de Luminy, F-13288 Marseille, France}
\affiliation{NIKHEF theory group, P.O. Box 41882, 1009 DB Amsterdam, The Netherlands}

\author{Hiromichi Nishimura}
\email{hnishimura@keio.jp}
\affiliation{Research and Education Center for Natural Sciences, Keio University}
\affiliation{RIKEN BNL Research Center, Brookhaven National Laboratory, Upton, NY 11973}

\author{Robert D. Pisarski}
\email{pisarski@bnl.gov}
\affiliation{Department of Physics, Brookhaven National Laboratory, Upton, NY 11973}

\author{Vladimir V. Skokov}
\email{vskokov@ncsu.edu}
\affiliation{Department of Physics, North Carolina State University, Raleigh, North Carolina 27695, USA}
\affiliation{RIKEN BNL Research Center, Brookhaven National Laboratory, Upton, NY 11973}

\begin{abstract}
We compute the term $\sim g^3$ in the free energy for a $SU(N)$ gauge theory with nonzero holonomy at nonzero temperature.  
If the holonomy is generated kinematically by the introduction of gauge invariant sources coupled to Polyakov loops, the contribution of charged (off-diagonal) gluons to the free energy at order $g^3$, ${\cal F}^{\left( 3\rm{:c.g.} \right)}$, is singular: ${\cal F}^{\left( 3\rm{:c.g.} \right)} \neq 0 $ without holonomy, but  ${\cal F}^{\left( 3 \rm{:c.g.} \right)}= 0$ when the holonomy is nonzero, even infinitesimally.
We show that the absence of the charged gluon contribution is required by gauge invariance alone and is therefore a universal feature.
\end{abstract}

\maketitle

\section{Introduction and summary}
\label{sec:Introduction}

Quantum chromodynamics (QCD) at a temperature $T$ is a subject of great beauty \cite{Kapusta:2006pm,Laine:2016hma}. 
It is of importance in the collisions of heavy ions at high energies and for the early universe.
At high temperature, the free energy can be computed perturbatively
to order $g^6$, where $g$ is the QCD coupling constant 
\cite{Collins:1974ky,Shuryak:1977ut,Kapusta:1979fh,Toimela:1982hv,Arnold:1994ps,Zhai:1995ac,Kajantie:2002wa}.  

In this paper we study how nonzero holonomy affects the free energy.
At tree level nonzero holonomy can be implemented by introducing a source coupled to the temporal component
of the gauge field, $A_4$.  Since this is not gauge invariant, though, while the associated free
energy  as a function of the source is gauge invariant to $\sim g^0$, it is gauge dependent to $\sim g^2$ \cite{Enqvist:1990ae}.

This can be avoided by using sources which are gauge invariant.  The natural choice are Polyakov loops~\cite{Fukushima:2017csk},
which are the order parameters for the deconfining phase transition in a $SU(N)$ gauge theory, without dynamical
quarks.  Denoting the sources which couple to Polyakov loops as $h$, to $\sim g^2$
the free energy is smoothly behaved as the holonomy vanishes, $h \rightarrow 0$
\cite{Bhattacharya:1990hk,Bhattacharya:1992qb,Belyaev:1991gh,KorthalsAltes:1993ca,Giovannangeli:2002uv,Giovannangeli:2004sg,altes2007,Dumitru:2013xna,Guo:2014zra,Guo:2018scp,KorthalsAltes:2020ryu}.
In this note we show that something unexpected happens to $\sim g^3$.  If $m_{\rm D} \sim g T$ is the
Debye mass at $h = 0$, then the free energy density from charged (off-diagonal) gluons has a singular behavior as $h \rightarrow 0$, 
\begin{equation}
\lim_{h \rightarrow 0}{\cal F}(h) - {\cal F}(0)
=  
\frac{N^2 -N}{12 \pi}  T m^3_D + \mathcal{O}(g^4) \; ,
 \label{MainResult}
\end{equation}
requiring that none of the eigenvalues of the Wilson line are degenerate for nonzero $h$.
We will discuss the meaning of this singular behavior at the end of Sec.~\ref{sec:Beyondg2}.
The details of the computation are presented in a longer work \cite{KorthalsAltes:2020ryu}.  Our purpose here is to ensure that
this admittedly odd result is not buried therein.  To anticipate these results, we stress that this quandary
arises when the holonomy is generated kinematically, by the introduction of
external sources.

\section{Constant background field}
\label{sec:ConstantBF}

In this section we first introduce a constant background gauge field and establish our notations. 
We then write down the perturbative expansion of the Polyakov loop around the constant background field up to order $g^2$, which is all we need to compute the free energy up to order $g^3$. 

\subsection{Conventions}
\label{sec:Conventions}

We take the Cartan basis for a $SU(N)$ gauge group.
There are $N^2-N$ off-diagonal generators $(T^{i j})_{k l} = 
\delta_{ik}  \delta_{jl}/\sqrt{2}$, with $i \neq j$;
$i,j,k,l = 1, \dots, N$ are indices
for the fundamental representation and $a,b,c = 1, \dots, N^2 -1 $ for the adjoint.
There are $N-1$ diagonal generators $T^d = \mbox{diag} \left( 1,1, \dots, -d, 0, \dots, 0 \right)/\sqrt{2d (d+1)}$,
with $d = 1,\ldots ,N-1$.
Defining $T^{\bar{a}} = \left( T^{a} \right)^{\dagger}$, $\overline{ij} = ji$ and $\bar{d} = d$, with
the generators normalized as $\mathrm{tr} (T^a T^b) = \delta^{a\bar{b}}/2$.
The structure constant is $i f^{abc} =  2 \mathrm{tr} \left( T^a \left[ T^b, T^c \right] \right)$.  We note that the structure constant is imaginary in this basis.  In this paper repeated indices
are summed over unless otherwise stated; in the following this should be clear from the context.

We use the background field method, where the gauge potential
$A_{\mu} = \bar{A}_{\mu}  + \mathcal{A}_{\mu}$, with $\bar{A}_{\mu}$ is the background field
and $\mathcal{A}_{\mu}$ the quantum field.
Non-trivial holonomy is introduced by taking the background field $\bar{A}_4$
to be diagonal and constant,
\begin{equation}
\left( \bar{A}_4 \right)_{ij}=  \frac{T}{g} \; \sum_{d=1}^{N-1} \; \Theta_d  \left( T^d \right)_{ij}
=
\frac{T}{g} \, \theta_i \, \delta_{ij}
.
\label{BackgroundField}
\end{equation}
Using $\left(T^d \right)_{ii} - \left(T^d\right)_{jj} = i f_{d, ij, ji}$, we obtain the relationship, $\theta_i - \theta_j = i f_{d,ij, ji} \Theta_d$.

We term the $N^2-N$ off-diagonal components as charged gluons (c.g.) and the
$N-1$ diagonal components as neutral gluons (n.g.).
In the $R_{\xi}$ gauge the gluon propagator is
\begin{equation}
D^{a b}_{\alpha \beta} 
=
\frac{\delta^{a \bar{b}}}{p^2_a}
\left( \delta_{\alpha \beta}  - \left( 1 - \xi \right) \frac{p^a_{\alpha} p^a_{\beta}}{p^2_a} \right) 
\label{propagator_Rxi}
\end{equation}
where $p^a_{\alpha} =( p^a_4, \mathbf{p} ) $ is the four-momentum,
with $\left| \mathbf{p} \right|  =p$ and $p^a_4 = ( 2 \pi  n + \theta_a )T$  for integral $n$.
The background holonomy $\theta_a = \theta_i - \theta_j$ for the charged gluons and $\theta_d = 0$ for the neutral.

\subsection{Perturbative expansion of the Polyakov loop}

The Polyakov loop is the trace of the temporal Wilson loop $\Omega(\mathbf{x})$, which measures the holonomy:
\begin{equation}
L_n (\mathbf{x})
=
\frac{1}{N} \, \mathrm{tr} \;
\Omega(\mathbf{x})^n
\;\;\;\;\;
\mbox{with}
\;\;\;\;\;
\Omega(\mathbf{x})
= 
\mathcal{P} \exp \left[ i g \int^{\beta}_0 d \tau A_4 (\tau, \mathbf{x})   \right] , 
\label{PolyakovLoop}
\end{equation}
where $\beta = 1/T$ and $\mathcal{P}$ denotes path ordering.  
The subscript $n$ indicates how many times the Wilson line wraps around in the temporal direction.  

We expand the Polyakov loop (\ref{PolyakovLoop}) in the
presence of the constant background field (\ref{BackgroundField}) to order $g^2$,
\begin{equation}
L_n (\mathbf{x}) = L^{\left( 0 \right)}_n +  L^{\left( 1 \right)}_n + 
 L^{\left( 2 \right)}_n  + \mathcal{O}(g^3).
\end{equation}
At leading order,
\begin{equation}
L^{\left( 0 \right)}_n 
=
\frac{1}{N } \mathrm{tr} \, e^{ i n \Theta} 
=
\frac{1}{N } \sum^{N}_{i=1} e^{i n \theta_i} .
\end{equation}
We denote a term with $\left( l \right)$ in the superscript as a term in the small-coupling expansion at order $g^{l}$.  The expansion up to order $g^2$ was first obtained in \cite{KorthalsAltes:1993ca},
but for completeness we rederive it here in a slightly different way.

The linear term is
\begin{eqnarray}
L^{\left( 1 \right)}_n  (\mathbf{x})
=
 \frac{i g}{N}  \mathrm{tr} \left[ e^{i g n \beta \bar{A}_4} T^d  \right] \int^{n \beta}_0 d\tau \mathcal{A}^d_4 ( \tau, \mathbf{x}),
\end{eqnarray}
after using the identity $e^{-i g \tau \bar{A}_4}  T^a e^{i g \tau \bar{A}_4}  = T^a e^{-i \tau T \theta^a} $.
Because the holonomy is a diagonal matrix,
only the neutral gluon $\mathcal{A}^d_4$ contributes.  
In momentum space,
\begin{equation}
\mathcal{A}_4 (\tau) =  \sum_{p_4} e^{- i p_4  \tau} \tilde{\mathcal{A}}_4 (p_4) 
\;\;\;\;\;
\mbox{with}
\;\;\;\;\;
\sum_{p_4} \equiv T \sum^{\infty}_{n_p = - \infty},
\end{equation}
so that
\begin{equation}
L^{\left( 1 \right)}_n
= 
g   \frac{ \partial L^{\left( 0 \right)}_n}{ \partial \Theta^d} \tilde{A}^d_4(0).
\label{L_1}
\end{equation}
Henceforth we drop the dependence on $\mathbf{x}$ for brevity.

To quadratic order
\begin{equation}
L^{\left( 2 \right)}_n 
=
-\frac{g^2}{N}  \mathrm{tr} \left[ e^{i g n \beta \bar{A}_4} T^b T^a \right]
\sum_{p_4, q_4}
 \tilde{ \mathcal{A}}^{b}_4(q_4)  \tilde{ \mathcal{A}}^{a}_4(p_4) \int^{n \beta}_0 d \tau_2 \int^{\tau_2}_{0} d \tau_1
e^{- i \left(q^b_4 \tau_2 + p^a_4 \tau_1 \right)} .
\label{L2_pspace}
\end{equation}
There are two types of contributions:
one from the charged gluons, $L^{\left( 2 : {\rm c.g.} \right)}_n$, when $a = ij = \bar{b}$;
the other is from neutral gluons, $L^{\left( 2 : {\rm n.g.} \right)}_n$, when $a=d$ and $b=d'$. 
For the contribution of neutral gluons,
the only nonzero contribution is from the zero mode with $p_4 = q_4 =0$.
In this case the time integral gives $n^2 \beta^2/2$, and so
\begin{equation}
L^{\left( 2: \rm{n.g.} \right)}_n 
=
\frac{g^2 }{2} \frac{\partial^2 L^{\left( 0 \right)}_n}{\partial \Theta^{d'} \partial \Theta^{d}}  
\tilde{ \mathcal{A}}^{d'}_4(0)  \tilde{ \mathcal{A}}^{d}_4(0) .
\label{L2_ng}
\end{equation}
For the charged gluon contribution, we use the identity,
$ \mathrm{tr} \left(e^{ig n \beta \bar{A}_4} T^{ji} T^{ij}   \right) =e^{i n \theta_j} /2$ with $i$ and $j$ fixed,
and perform the time integral to obtain
\begin{equation}
L^{\left( 2 {\rm: c.g.} \right)}_n 
=
-\frac{g^2}{2 N}  
\sum_{p_4, q_4}
\tilde{ \mathcal{A}}^{ji}_4(q_4)  \tilde{ \mathcal{A}}^{ij}_4(p_4) 
\left[ \frac{i n e^{in \theta_j}}{p^{ij}_4 T} \delta (p_4 + q_4)
-
\frac{ e^{in \theta_j} -  e^{in \theta_i}}{p^{ij}_4 q^{ji}_4}  \right] .
\end{equation}
The second term in the bracket is odd under $p_4 \leftrightarrow q_4$ and $i \leftrightarrow j$,
and thus vanishes. Therefore the only nonzero contributions are from the first term, with $q^b_{4} = - p^a_{4}$.  
Performing the $q_4$ sum and using the identity,
\begin{eqnarray}
\frac{in}{2N} \left( e^{in \theta_i} - e^{in \theta_j}   \right)
=
\frac{i n}{N} 
 \mathrm{tr} \left( e^{i g n \beta \bar{A}_4} \left[T^{ij}, T^{ji} \right] \right)
=
i f^{d, ij, ji} \frac{\partial L^{\left( 0 \right)}_n}{\partial \Theta^d},
\label{identity_1}
\end{eqnarray}
we obtain
\begin{equation}
L^{\left( 2: {\rm c.g.} \right)}_n 
=
\frac{g^2 }{2} 
i f^{d, ij, ji}
\frac{\partial L^{\left( 0 \right)}_n}{\partial \Theta^d}
\sum_{p_4}  \frac{1}{p^{ij}_4}
 \tilde{ \mathcal{A}}^{ji}_{4}(-p_4)  
\tilde{ \mathcal{A}}^{ij}_{4}(p_4)  .
\label{L2_cg}
\end{equation}
The total contribution to the quadratic term in the expansion of the Polyakov loop is the sum of the two terms, $L^{\left( 2 \right)}_n = L^{\left( 2: {\rm n.g.} \right)}_n  + L^{\left( 2: {\rm c.g.} \right)}_n $, Eqs.~(\ref{L2_ng}) and (\ref{L2_cg}).  

\section{Effective potential for the holonomy}
\label{sec:Deformation}

\subsection{Deformed theory}

To probe nontrivial holonomy, we add a deformation to Yang-Mills action $S_{\rm YM}$,
\begin{equation}
Z(h) = \int DA \exp \left[-S_{\rm YM}(A) + \Delta S(h, A_4)   \right] 
\;\;\;\;\; \mbox{with} \;\;\;\;\; 
\Delta S=  T^3  \int d^3 x \, h \, U( L_n ) \, .
\label{Z_h}
\end{equation}
For convenience we choose both $h$ and $U$ to be dimensionless. Here we use a single source $h$, but the extension to multiple sources is straightforward.  We require that the deformation $ \Delta S$ shifts the expectation value of the Polyakov loop smoothly.

This theory can be interpreted in various ways.  
In the semiclassical language, $S_{\rm YM}$ and $\Delta S$ can be seen as a perturbative and non-perturbative part of the YM theory, respectively.  In phenomenology, the term $\Delta S$ is an additional term, which is constructed by hand to model
the deconfining phase transition.  A natural choice is to use a ``double-trace'' deformation,
\begin{equation}
h U = h \sum^{\infty}_{n=1} \frac{1}{n^2} \left| L_n \right|^2.
\label{U_B2}
\end{equation}
The eigenvalues $\theta_i$ become non-degenerate once the source is turned on. 
This form was considered in phenomenological models for the deconfining phase transition \cite{Meisinger:2001cq,Dumitru:2010mj} and in confining gauge theories on $R^3 \times S^1$ \cite{Myers:2007vc,Unsal:2008ch}. The latter subject has developed significantly in recent years in the context of the large-$N$ orbifold equivalence and resurgence \cite{Dunne:2015eaa}. 
A few specific forms of the deformation, including Eq.~(\ref{U_B2}), are considered in \cite{KorthalsAltes:2020ryu}.

We compute the effective potential for the holonomy for an arbitrary deformation.
Using the constant background field $\bar{A}_{\mu}= \delta_{\mu 4} \Theta / \left(g \beta \right)$
as given in Eq.~(\ref{BackgroundField}), the effective potential for the holonomy is
\begin{equation}
\exp \left[- \beta \mathcal{V} V_{\rm eff} (\Theta, h) \right] 
= 
\int_{\rm 1PI}  
 D \mathcal{A} \exp \left[-S_{\rm YM}(\mathcal{A}+ \bar{A}) +  T^3  \int d^3x \, h U(L_n (\mathcal{A}_4 + \bar{A}_4))   \right] ,
 \label{V_eff}
\end{equation}
where $\mathcal{V}$ is the spatial volume.  This consists of two parts,
\begin{equation}
V_{\rm eff} (\Theta, h) = V_{\rm pert} (\Theta) + \Delta  V (\Theta, h),
\end{equation}
where the first term is the usual perturbative contribution and the second is due to the deformation.   Both
are one particle irreducible (1PI).

In the deformed theory expanding $U$ generates additional terms.  To order $g^2$,
\begin{equation}
U
=
U^{\left( 0 \right)} 
+\frac{ \partial U^{\left( 0 \right)} }{\partial L^{\left( 0 \right)}_n} L^{\left( 1 \right)}_n  
+
\frac{ \partial U^{\left( 0 \right)} }{\partial L^{\left( 0 \right)}_n} L^{\left( 2\right)}_n  
+ 
\frac{1}{2} \frac{ \partial^2 U^{\left( 0 \right)}  }{\partial L^{\left( 0 \right)}_m \partial L^{\left( 0 \right)}_n} L^{\left( 1\right)}_m L^{\left( 1\right)}_n 
 + \mathcal{O}(g^3)
 \label{U_expansion}
\end{equation}
where $U^{\left( 0 \right)}  \equiv U(L^{\left( 0 \right)}) $ is of order $g^0$.
The repeated indices are summed over from $n=1$ to $N-1$.  The second term $\sim g$ is a tadpole,
which by Eq.~(\ref{L_1}) is irrelevant to a potential which is one particle irreducible.

The last two terms in Eq.~(\ref{U_expansion}) are $\sim g^2$, denoted as $U^{\left(2 \right)}$.
We treat them as interaction terms, and show in the next section that they modify the gluon self-energy.
Using Eqs.~(\ref{L2_ng}) and (\ref{L2_cg}), we obtain $U^{ \left( 2 \right) }  = U^{ \left( 2: \rm{n.g.} \right)} + U^{ \left( 2: \rm{c.g.} \right)}$ with
\begin{eqnarray}
U^{\left( 2:{\rm n.g.} \right)} 
&=&
\frac{g^2 }{2} 
\frac{\partial^2 U^{\left( 0 \right)} }{\partial \Theta^{d'} \partial \Theta^{d}}
\tilde{ \mathcal{A}}^{d'}_4(0)  \tilde{ \mathcal{A}}^{d}_4(0) 
\label{U_g2_ng}
\\
U^{\left( 2:{\rm c.g.} \right)} 
&=&
\frac{g^2 }{2} 
i f^{d, ij, ji} \frac{\partial U^{\left( 0 \right)}}{\partial \Theta^d}
\sum_{p_4}  \frac{1}{p^{ij}_4}
 \tilde{ \mathcal{A}}^{ji}_{4}(-p_4)  
\tilde{ \mathcal{A}}^{ij}_{4}(p_4) 
\label{U_g2_cg}
\end{eqnarray}
where
$U^{ \left( 2: \rm{n.g.} \right)}$  and  $U^{ \left( 2: \rm{c.g.} \right)}$ are the quadratic terms for the neutral and charged gluons from the expansion of $U$, respectively. We note that the former has only the zero Matsubara mode.  
These two terms together with the usual three and four-gluon vertices from $S_{\rm YM}$ are all we need to construct the 1PI diagrams up to order $g^3$.

\subsection{Free energy}

The saddle point of the effective potential is determined by the equation of motion,
\begin{equation}
\left. 
\frac{\partial }{\partial \Theta_d} V_{\rm eff} (\Theta, h) \right|_{\Theta_d = \Theta^{*}_{d}}
=
0
\label{EoM}
\end{equation}
which can be computed order by order.  We parametrize the saddle point as $\Theta^{*}(h)
=
\Theta^{\left( 0 \right)} (h)
+
\Theta^{\left( 2 \right)} (h)
+
\mathcal{O}(g^3)$
up to order $g^2$.
The free energy density $\mathcal{F}$ is given by the saddle point of the effective potential,
\begin{eqnarray}
\mathcal{F}(h)
=
V_{\rm eff } (\Theta^{*}, h)
=
V_{\rm eff } (\Theta^{\left( 0 \right)}, h) 
+ \frac{1}{2} \Theta^{\left( 2 \right)}_d \Theta^{\left( 2 \right)}_{d'} \frac{ \partial^2 }{\partial \Theta^{\left( 0 \right)}_d \partial \Theta^{\left( 0 \right)}_{d'}} V_{\rm eff } (\Theta^{\left( 0 \right)}, h) + \mathcal{O}(g^5) ,
\label{FreeEnergy}
\end{eqnarray}
where the linear term vanishes by Eq. (\ref{EoM}).

At leading order the effective potential
\begin{equation}
\beta^4 V^{\left( 0 \right)}_{\rm eff} 
=
\frac{2 \pi^2 }{3} \sum_{a} B_{4} \left( \frac{\theta_a}{2 \pi }\right) 
 - h U^{\left( 0 \right)} .
 \label{V_g0}
\end{equation}
The first term is familiar at one loop order \cite{Gross:1980br,Weiss:1980rj}, with
$B_k(x)$ is the $k$-th Bernoulli polynomial, defined for $0 \leq x \leq 1$.  
Due to the periodicity of the holonomy, the argument of the Bernoulli polynomial is understood to be $\theta_a$ mod $2 \pi$.
At this order, the saddle point is
\begin{equation}
h \frac{\partial U^{\left( 0 \right)}}{\partial \Theta_d}
=
i \frac{4 \pi }{3} f_{da\bar{a}} B_{3} \left( \frac{\theta_a}{2 \pi }\right) .
\label{EoM_g0}
\end{equation}
We assume that the deformation $U$ is such that the saddle point is nonzero when $h\neq 0$.

The next-leading contribution is of order $g^2$,
$V^{\left(2 \right)}_{\rm eff} = V^{\left( 2 \right)}_{\rm pert} + \Delta V^{\left( 2 \right)}$. 
The perturbative potential $V^{\left( 2 \right)}_{\rm pert} $
has been computed previously in $R_\xi$ gauge
\cite{Anishetty:1981tk,Belyaev:1989bj,Enqvist:1990ae,Skalozub:1992un,KorthalsAltes:1993ca,Dumitru:2013xna}. 
There is an additional contribution, $\Delta V^{\left( 2 \right)}$,
from the Wick contraction of $T^3 \int d^3 x \, h U^{\left( 2 \right)}$.
The contribution of neutral gluons,
$U^{\left( 2:{\rm n.g.} \right)} $ in Eq.~(\ref{U_g2_ng}),
gives $\sim g^2 \int_{\mathbf{p}} 1/ p^2$, vanishes with dimensional regularization.
The contribution from charged gluons, $U^{\left( 2:{\rm c.g.} \right)} $ in Eq.~(\ref{U_g2_cg}),
is nonzero:
\begin{eqnarray}
\beta^4 \Delta V^{\left( 2 \right)} (\Theta, h)
&=&
 - ig^2  \frac{\left( 3- \xi \right)}{8 \pi} 
f_{da \bar{a}} 
B_1 \left( \frac{\theta_a}{2 \pi} \right)
h \frac{\partial U^{\left( 0 \right)}}{\partial \Theta^d} .
\label{DeltaV_g2}
\end{eqnarray}
It appears that the free energy as a function of holonomy at order $g^2$ is  gauge dependent due to the presence of $\xi$ in $\Delta V^{\left(2 \right)} $. 
This $\xi$-dependent term, however, is cancelled by a
$\xi$-dependent term from $V^{\left( 2 \right)}_{\rm pert}$.  Combining the two,
\begin{equation}
\mathcal{F}^{\left( 2 \right)}
=
g^2 T^4
\sum_{ijk}\left[
\frac{1}{4}
B_2 \left( \frac{\theta^{\left( 0 \right)}_{ij}}{2 \pi} \right) B_2 \left( \frac{\theta^{\left( 0 \right)}_{ik}}{2 \pi} \right) 
- 
\frac{1 }{4 N^2} B^2_2(0)
-
\frac{2}{3} B_1 \left( \frac{\theta^{\left( 0 \right)}_{ij}}{2 \pi} \right) B_3 \left( \frac{\theta^{\left( 0 \right)}_{ik}}{2 \pi} \right) 
\right] \; ,
\label{F_g2_1}
\end{equation}
where $\theta^{\left( 0 \right)}_a = i f_{d a \bar{a}}\Theta^{\left( 0 \right)}_d $.
Thus to $\sim g^2$ the free energy is independent of the gauge parameter for any value of $h$, as expected for sources
which are gauge invariant.
This result agrees with Ref. \cite{Dumitru:2013xna},
which uses the constraint field method to construct the effective potential.    
In the limit $h \rightarrow 0$,
$\mathcal{F}^{\left( 2 \right)} \rightarrow Ng^2 \left(N^2 - 1 \right) T^4 /144$ \cite{Shuryak:1977ut}.   

Thus to $\sim g^2$ the free energy in the deformed theory is smoothly connected to that for
$h = 0$ as $h \rightarrow 0$.
We now show that this continuity is lost as we go to higher order.

\section{Beyond order $g^2$}
\label{sec:Beyondg2}

\subsection{Self-energy at order $g^2$}

We begin by computing the gluon self energy to $\sim g^2$ in the static limit, which is necessary for the free energy
$\sim g^3$ in the next subsection.  For nonzero deformation in Eq. (\ref{Z_h}), the self-energy
consists to two parts,
\begin{equation}
\Pi^{ab}_{\alpha \beta} 
 = 
\left( \Pi_{\rm pert} \right)^{ab}_{\alpha \beta} 
+
\left(\Delta  \Pi \right)^{ab}_{\alpha \beta} ,
\label{Pi}
\end{equation}
where $\Pi_{\rm pert} $ is the usual one-loop self-energy in a background field,
and $\Delta \Pi$ is the additional contribution due to the deformation.
The complete expression for $\Pi_{\rm pert}$ for nonzero holonomy
in Feynman gauge is given in \cite{Nishimura:2018wla}.  Here we only show the results needed for our discussion.

Because of the conservation of color charge, the charged gluons with holonomy $\theta_{ij}$
do not mix with other gluons.  The self-energy for the charged gluon, written as  $\Pi^{ij,ji}_{\mu \nu}$,
is thus diagonal in the color space.  On the other hand,  neutral gluons mix with one another,
so the associated self-energy,  $\Pi^{dd'}_{\mu \nu}$, is not diagonal.  

The gluon self-energy is only transverse at zero temperature.  At nonzero temperature without holonomy \cite{Elze:1987rh},
%
 \begin{equation}
p_{\mu} \left( \Pi_{\rm pert} \right)_{\mu \nu}(p) 
= 
(1-\xi) g^2 N T_{\nu \rho}(p) 
\int_q  \frac{q \cdot r}{q^4 r^2} q_{\rho} .
\label{Pi_longitudinal}
\end{equation}
Here $p,q,r$ with $p+q+r=0$ is the four-momentum without holonomy, and $T_{\rho\nu}(p)=p^2\delta_{\rho\nu}-p_\rho p_\nu$.

In the presence of the background field, 
there is an additional term on the RHS of Eq.~(\ref{Pi_longitudinal}). 
The self-energy for the charged gluons picks up an extra part due entirely to the background \cite{KorthalsAltes:1993ca,Hidaka:2009hs,Nishimura:2018wla},
\begin{eqnarray}
p^{ij}_{\mu} \left( \Pi_{\rm pert} \right)^{ij, ji}_{\mu \nu} 
&=&
\frac{4 \pi}{3} g^2 T^3 \sum^N_{k=1} \left[B_3\left(\frac{\theta_{ik}}{2\pi}\right) - B_3\left(\frac{\theta_{jk}}{2\pi}\right) \right]\delta_{4 \nu}
\label{KA_cg}
\end{eqnarray}
in the Feynman gauge, $\xi =1$.  

We now take the static limit, $p_4 =0$ and $ \mathbf{p}  \rightarrow 0$.  
In this limit terms as in (\ref{Pi_longitudinal}) play no role anymore.

$\Pi_{\rm pert}$ for the neutral gluon is given in Refs. \cite{Giovannangeli:2002uv,Nishimura:2018wla,KorthalsAltes:2020ryu}:
\begin{eqnarray}
\left(\Pi_{\rm pert} \right)^{d d'}_{44} (0,  \mathbf{0}) 
&=& 
2 g^2 T^2 f_{d b c} f_{d' \bar{b} \bar{c}} B_2 \left( \frac{\theta_b}{2 \pi} \right) .
\label{Pi_YM_ng_SL}
\end{eqnarray}
For the charged gluons, we have 
\begin{eqnarray}
\left(\Pi_{\rm pert} \right)^{ij, ji}_{44} (\theta_{ij},  \mathbf{0})
&=&
    \frac{4 \pi}{3} g^2 T^2 \sum^N_{k=1} \frac{1}{\theta_{ij}} \; \left(
    B_3\left(\frac{\theta_{ik}}{2\pi}\right) - B_3\left(\frac{\theta_{jk}}{2\pi}\right) \right) \; .
\label{Pi_cg_SL}
\end{eqnarray}
This equality follows from Eqs.~(\ref{Pi_longitudinal}) and (\ref{KA_cg}), where we write $p^{ij}_{\mu} \left( \Pi_{\rm pert} \right)^{ij, ji}_{\mu 4} = p^{ij}_{4} \left( \Pi_{\rm pert} \right)^{ij, ji}_{4 4} +  p^{ij}_{x} \left( \Pi_{\rm pert} \right)^{ij, ji}_{x 4}$. We can then show that $\left(\Pi_{\rm pert} \right)^{ij, ji}_{x 4}$ is a smooth function of $p$, and  the RHS of Eq.~(\ref{Pi_longitudinal}) with the holonomy vanishes in the static limit. The self-energy at this order is therefore gauge invariant in the static limit. 
In the limit of vanishing holonomy, we recover the usual self-energy in pure Yang-Mills theory without holonomy,
\begin{equation}
\lim_{\theta \rightarrow 0} \left(\Pi_{\rm pert} \right)^{ab}_{44} (\theta_{a},  \mathbf{0})
=
\frac{1}{3} N g^2 T^2 
\delta^{a \bar{b}}
= 
m^2_D \delta^{a \bar{b}},
\end{equation}
where $m_D$ is the Debye mass.  

In the usual background field method where the external source is coupled to $A_4$,
the self-energy consists only of $\Pi_{\rm pert}$.
The non-zero contribution of charged gluons (\ref{Pi_cg_SL}) then gives rise to unphysical results at order $g^3$ when $\theta \neq 0$, as we discuss in the next subsection.

There is, however, an additional contribution to the
self-energy in the presence of the deformation. This comes from the expansion of $U$ (\ref{U_expansion}).
Neutral gluons contribute to the gluon self energy as
\begin{equation}
 \left( \Delta \Pi \right)^{dd'}_{\mu \nu} 
 =
-g^2 T^2 h  \frac{\partial^2 U^{\left( 0 \right)}}{\partial \Theta^{d'} \partial \Theta^{d}} \delta (p_4) \delta_{4 \mu} \delta_{4 \nu}.
 \label{DeltaPi_ng}
\end{equation}
This is transverse because it only contributes for the zero mode. 
On the other hand, using Eq.~(\ref{U_g2_cg})
\begin{equation}
 \left( \Delta \Pi \right)^{ij, ji}_{\mu \nu} 
 =
- g^2 T^3 h  \frac{\partial U^{\left( 0 \right)} }{\partial \Theta^d}   i f^{d, ij, ji}  \frac{1}{p^{ij}_4}
 \delta_{4 \mu}  \delta_{4 \nu} 
 \label{DeltaPi_cg}
\end{equation}
for the charged gluons. 
We note that these contributions are independent of the gauge parameter at the leading order.

In the static limit, $\Delta \Pi$ for the neutral gluons depends on the
specific form of $U$ according to Eq.~(\ref{DeltaPi_ng}).  Nevertheless it vanishes
as $h \rightarrow 0$ regardless of $U$. Therefore the total
self-energy for the neutral gluons becomes the usual Debye mass as the holonomy vanishes.  
On the other hand, $\Delta \Pi$  for the charged gluons does not vanish in the limit $p_4 =0$ and
$\mathbf{p} \rightarrow 0$. According to Eq.~(\ref{DeltaPi_cg}), we have
\begin{equation}
\left( \Delta \Pi \right)^{ij, ji}_{44} (\theta_{ij},  \mathbf{0}) = -\left(\Pi_{\rm pert} \right)^{ij, ji}_{44} (\theta_{ij},  \mathbf{0}) 
\label{DeltaPi_cg_44}
\end{equation}
at the saddle point (\ref{EoM_g0}). Therefore the total self-energy for
the charged gluons in this limit vanishes identically for any value of nonzero $\theta$. 
In particular, as we take $\theta \rightarrow 0$, we have
$\left( \Pi_{\rm pert} \right)^{ij,ji}_{44} \rightarrow m^2_D$
and $\left( \Delta \Pi \right)^{ij,ji}_{44} \rightarrow -m^2_D$, so we do not recover
the usual result without holonomy. This is the source of the singular behavior in Eq.~(\ref{MainResult}).

\subsection{Free energy at order $g^3$}

For weak holonomy, the leading correction to the free energy at $\sim g^2$ is $\sim g^3$.
This contribution comes from the sum of the set of the so-called ring diagrams:
\begin{equation}
\mathcal{F}^{\left( \rm{ring} \right)}
=
-\frac{1}{2} \sum^{\infty}_{k=2}  \frac{1}{k} \sum_{p_4} \int \frac{d^3 p}{\left( 2\pi \right)^3} 
\mathrm{tr} 
\left[ 
- \Pi^{ab}_{\alpha \beta}  D^{bc}_{\beta \gamma}
 \right]^k
\end{equation}
where $k$ is the number of the ring, i.e.~the self-energy, and the prefactor $1/k$ accounts for the symmetry factor.  
The propagator $D$ in the $R_\xi$ gauge is given in Eq.~(\ref{propagator_Rxi}).  

Let us first review the usual case without holonomy, $h = \theta =0$, and the self-energy $\Pi = \Pi_{\rm pert} $. 
Each ring diagram is infrared divergent in
the static limit, $p_4 =0$ and $\mathbf{p} \rightarrow 0$.
In this limit the only nonzero term is the fourth component of the self-energy,
which is the usual Debye mass squared,
$\lim_{\mathbf{p}  \rightarrow 0} \Pi^{44}_{ab} (p_4 = 0, \mathbf{p} ) = m^2_D \delta_{a \bar{b}}$.
The infrared divergence is cured by the Debye screening, and this gives rise to the $g^3$ contribution to the free energy \cite{Kapusta:1979fh},
\begin{equation}
\mathcal{F}^{\left(3 \right)} \left( h =0 \right)
=
\frac{1}{2} T \sum_a \int \frac{d^3 p}{\left( 2 \pi \right)^3} \log \left( 1 + \frac{m^2_D}{p^2} \right)
=
- \frac{N^2 - 1}{12 \pi} T m^3_D 
\label{F_g3_pQCD}
\end{equation}
with dimensional regularization, 
where $p = \left| \mathbf{p} \right|$.
We note that both $N^2-N$ charged gluons and
the $N-1$ neutral gluons contribute to $\sim g^3$,
giving the prefactor of $N^2-1$.  It is important that the resummation consists only of the zero Matsubara mode,
so that the gauge parameter $\xi$ in the propagator drops out.

We now turn on the external source $h$ and consider nontrivial holonomy $\theta \neq 0$. 
As discussed in the previous subsection, the self-energies for the charged and neutral gluons do not mix,
so the ring resummation can be written as a sum of two contributions from the charged and neutral gluons.  

The contribution of neutral gluons is similar to the case without holonomy.
The zero-mode of the ring diagram $p_4 = 0$ for $k \geq 2$ is again infrared
divergent for any value of $\theta$.
Denoting $\Lambda^{d}_4$ as the eigenvalue of the self-energy in the static limit,
\begin{equation}
\mathcal{F}^{\left( 3:{\rm n.g.} \right)} (h)
=
\frac{1}{2} \sum^{N-1}_{d=1} \int \frac{d^3 p}{\left( 2 \pi \right)^3}  \left[ \ln \left(1 + \frac{\Lambda^{d}_4}{p^2} \right) - \frac{\Lambda^d_4}{p^2} \right]
=
-\frac{1}{12 \pi} T \sum^{N-1}_{d=1} \left( \Lambda^{d}_4 \right)^{3/2} .
\label{F_g3_ng} 
\end{equation}
As we take $h \rightarrow 0$, $\Delta \Pi \rightarrow 0$ and thus
$\Lambda^d_4 \rightarrow m^2_D$, and the usual contribution of neutral gluons is recovered.

The charged-gluon contribution to the ring diagram  is more subtle.
Unlike previous cases, the $\xi$ dependent term in the propagator for the
zero Matsubara mode does not vanish due to the nontrivial factor of $\theta$.
This implies that in the usual background field method where the external source
is coupled to $A_4$, the ring diagram becomes dependent upon $\xi$.
In our case, this problem should not occur because the deformation is gauge invariant. Indeed the additional contribution $\Delta \Pi$ due to the deformation gives a gauge-invariant result, but in a striking way: because the total self-energy in the static limit vanishes, the IR divergence is absent at this order, and so no resummation is needed for the charged gluons.  
Thus
\begin{equation}
\mathcal{F}^{\left( 3:{\rm c.g.} \right)} (h)= 0
\label{F_g3_cg}
\end{equation}
as long as $h$ is nonzero.
For the charged gluons, the infrared problem at order $g^3$ is cured not by the screening, but by the cancellation of the infrared divergence by the additional contribution $\Delta \Pi$. 

The total free energy at order $g^3$ is the sum of Eqs.~(\ref{F_g3_ng}) and (\ref{F_g3_cg}). 
The $g^3$ contribution comes from the ring diagrams of neutral gluons only, and it is therefore $1/N$ suppressed at large $N$. 
This holds true for any arbitrary small value of $h$.  
Combining Eqs. (\ref{F_g3_pQCD})-(\ref{F_g3_cg}), we obtain Eq.~(\ref{MainResult}). 

We would like to point out that this singular behavior, 
$\mathcal{F} (h \rightarrow 0) \neq \mathcal{F} (h=0)$, does not signal any phase transition. The free energy in the deformed theory is continuous for any nonzero value of $h$. What Eq.~(\ref{MainResult}) tells us is that the deformed theory (\ref{Z_h}) (even with arbitrary small deformation) is not continuously connected to pure Yang-Mills theory in perturbation theory.  If we first take the limit $h \rightarrow 0$ in Eq.~(\ref{Z_h}) and then perform a perturbative expansion, then we get the usual result in pure Yang-Mills theory.  On the other hand, if we first perform a perturbative expansion of the deformed theory (\ref{Z_h}), then we obtain a different result at order $g^3$ as we take $h \rightarrow 0$. 

We have explicitly seen that this singular behavior stems from Eq.~(\ref{DeltaPi_cg_44}): The deformation gives rise to the additional contribution to the one-loop self energy, which is absent in pure Yang-Mills theory. We have shown that this contribution is essential to keep the theory gauge invariant, and it does not vanish even in the limit $h \rightarrow 0$.  This modifies the total one-loop self-energy for the charged gluons, and it in turn changes the free energy at order $g^3$, as well as other physical quantities that depend on the one-loop self-energy.  We thus conclude that the singular behavior in perturbation theory (\ref{MainResult}) is inevitable once we introduce the holonomy by kinematically adding any gauge-invariant deformation (\ref{Z_h}) to pure Yang-Mills theory.

\section{Conclusions}
\label{sec:Conclusions}

We have considered the free energy in the presence of the nontrivial holonomy up to order $g^3$ in perturbation theory.
In order to probe the nontrivial holonomy in an explicitly gauge invariant way, we introduce an
external source $h$ that couples to a gauge invariant term $U$ (\ref{Z_h}).
We have shown that the free energy is indeed gauge invariant,
and it is smoothly connected to the result without holonomy to $\sim g^2$.

The nontrivial holonomy, however, changes the infrared behavior of the theory.
In the deformed theory, we have additional contribution to the self-energy at order $g^2$, and the total self-energy for the charged gluons vanishes in the static limit. 
This leads us to conclude that there is no $g^3$ contribution from the charged gluons in the deformed theory.
The deformed theory is singular in a sense that we do not recover the usual
result as $h \rightarrow 0$. 

The results obtained here are valid for a wide class of observables, like the Polyakov loop probability distribution
used on the lattice.   We investigate this further in Ref. \cite{KorthalsAltes:2020ryu}.

\acknowledgements

We thank  Bengt Friman, Yoshimasa Hidaka, Krzysztof Redlich, and Michael Strickland for useful discussions.  
C.P.K.A.  acknowledges hospitality extended to him by the Physics Department of  Brookhaven National Laboratory.
H.N. was supported by the Special Postdoctoral Researchers program of RIKEN and the Japan Society for the Promotion of Science (JSPS) Grant-in-Aid for Scientific Research (KAKENHI) Grant Number 18H01217.
R.D.P. is funded by the U.S. Department of Energy for support under contract DE-SC0012704.
V.S. is supported by the US Department of
Energy grant DE-SC0020081. 
V.S. thanks the ExtreMe Matter Institute EMMI (GSI
Helmholtzzentrum f\"ur Schwerionenforschung, Darmstadt, Germany) for partial support and
hospitality.

\bibliography{g3_short}
\newpage

\end{document}